\documentclass[a4paper,11pt,amsmath,amssymb]{article} 
\usepackage{jheppub}
\usepackage{amsmath} 
\usepackage{array,relsize,float}
\usepackage{pstricks}
\usepackage{color} 
\usepackage{amssymb}
\usepackage{slashed}
\usepackage{tikz-cd}
\usepackage{graphicx} 
\usepackage{epsfig}
\usepackage{multicol} 
\usepackage{slashed} 
\usepackage{hyperref}
\usepackage{pdfpages}
\usepackage{array}
\usepackage{tikz-cd}
\usepackage{amsfonts,layout,appendix,subfigure}
\allowdisplaybreaks
\input{epsf} 

\renewcommand{\thefigure}{\arabic{figure}}

\def\beq{\begin{equation}} \def\eeq{\end{equation}}
\def\beqn{\begin{eqnarray}} \def\eeqn{\end{eqnarray}}
 \def\to{\rightarrow}
\def\nn{\nonumber}

\def\ln#1{\mathrm{log}\left(#1\right)}

\def\beq{\begin{equation}}
\def\eeq{\end{equation}}
\def\bea{\begin{eqnarray}}
\def\eea{\end{eqnarray}}
\def\beqn{\begin{eqnarray}} \def\eeqn{\end{eqnarray}}
\def\beeq{\begin{eqnarray}}
\def\eeeq{\end{eqnarray}}

\def\nn{\nonumber}

\def\ln#1{\mathrm{log}\left(#1\right)}

\newcommand\as{\alpha_{\mathrm{S}}}

\def\as{\alpha_{\rm S}}

\def\muRsq{\mu_R^2}
\def\muFsq{\mu_F^2}

\newcommand{\valencia}{Instituto de F\'{\i}sica Corpuscular, Universitat de Val\`{e}ncia -- Consejo Superior de Investigaciones Cient\'{\i}ficas, Parc Cient\'{\i}fic, E-46980 Paterna, Valencia, Spain.}
\newcommand{\europea}{Escuela de Ciencias, Ingenier\'ia y Diseño, Universidad Europea de Valencia, Paseo de la Alameda 7, 46010 Valencia, Spain.}
\newcommand{\milano}{Tif Lab, Dipartimento di Fisica, Universit\'a di Milano and INFN, Sezione di Milano, Via Celoria 16, I-20133 Milan, Italy.}
\newcommand{\salamanca}{Departamento de F\'isica Fundamental, Universidad de Salamanca, E-37008 Salamanca, Spain.}

\begin{document}

\title{Combining QED and QCD transverse-momentum resummation for \boldmath$W$ and \boldmath$Z$ boson production at
hadron colliders}
\author[a]{Andrea Autieri,}
\author[a]{Leandro Cieri,}
\author[b]{Giancarlo Ferrera}
\author[c,d]{and German~F. R.~Sborlini} 
\affiliation[a]{\valencia}
\affiliation[b]{\milano}
\affiliation[c]{\salamanca}
\affiliation[d]{\europea}

\emailAdd{andrea.autieri@ific.uv.es}
\emailAdd{leandro.cieri@ific.uv.es}
\emailAdd{giancarlo.ferrera@mi.infn.it}
\emailAdd{german.sborlini@usal.es}

\preprint{IFIC/23-06, FTUV-22-1126.2949}

\abstract{In this article, we consider the transverse momentum ($q_T$) distribution of $W$ and $Z$ bosons produced in hadronic collisions. We combine the $q_T$ resummation for QED and QCD radiation including the QED soft emissions from the $W$ boson in the final state. In particular, we perform the resummation of enhanced logarithmic contributions due to soft and collinear emissions at next-to-leading accuracy in QED, leading-order accuracy for mixed QED-QCD and next-to-next-to-leading accuracy in QCD. In the small-$q_T$ region we consistently include in our results the next-to-next-to-leading order (i.e.\ two loops) QCD corrections and the next-to-leading order  (i.e.\ one loop) electroweak corrections. The matching with the fixed-order calculation at large $q_T$ has been performed at next-to-leading order in QCD (i.e.\ at $\mathcal{O}(\alpha_S^2)$) and at leading order in QED. We show numerical results for $W$ and $Z$ production at the Tevatron and the LHC. Finally, we consider the effect of combined QCD and QED resummation for the ratio of $W$ and $Z$ $q_T$ distributions, and we study the impact of the QED corrections providing an estimate of the corresponding perturbative uncertainties.}

\setcounter{page}{1}
\maketitle

\section{Introduction}
\label{sec:introduction}
The Drell-Yan mechanism
\cite{Drell:1970wh,Christenson:1970um},
i.e the production of 
high-invariant mass lepton pairs through the decay of an electroweak (EW) boson ($\gamma^{*}$, $Z$, $W$), is one of
the key-process at present and future hadron colliders (Tevatron \cite{Holmes:2011ey}, LHC \cite{Evans:2008zzb} and FCC-hh \cite{FCC:2018byv, FCC:2018vvp, FCC:2018bvk}) for validating Standard Model (SM) and searching new physics signals owing to Beyond Standard Model (BSM) effects.
\\
Historically, this mechanism was the first process in which the ideas of parton model and factorization, initially developed for the $\textit{deep inelastic lepton-hadron-scattering}$ (DIS) \cite{Bjorken:1968dy, nla.cat-vn431334, Altarelli:1977zs}, were applied in the context of hard-scattering hadron-hadron collisions. Nowadays, the achievement of a great level of accuracy for Drell-Yan observables is desirable for various reasons.
\\
Firstly, due to the the high production rates and a clear experimental signature given by a leptonic final state, the hadro-production of electroweak bosons is important for detector calibration, luminosity monitor and to probe underlying events \cite{CMS:2019raw,CMS:2017gbl,CMS:2021xjt,Khoze:2000db,CMS-PAS-FSQ-16-008,Grafstrom:2015foa}, besides representing a significant background for other SM processes \cite{ATLAS:2017cez,CMS:2017zyp} and BSM signals such as supersymmetric particles, new gauge bosons, heavy resonances, among other examples \cite{CMS:2021ctt,Leike:1998wr,Langacker:2008yv,CMS:2016ifc}.
\\
Furthermore, the Drell-Yan process gives us: \emph{(i)} strong tests of perturbative QCD (pQCD), of lepton universality and more in general of the Standard Model \cite{CMS:2018mdl,CMS:2022uul,ATLAS:2016nqi}; \emph{(ii)} stringent information and constraints on \textit{parton distribution functions} (PDFs) of the colliding hadrons \cite{CMS:2016qqr,Accomando:2017scx,Basso:2015lua}, and \emph{(iii)} precise measurements of the electroweak mixing angle \cite{CMS:2011utm}, the $W$ boson decay-width \cite{Camarda:2016twt} and mass \cite{ATLAS:2017rzl,CDF:2022hxs}. Along this line, hints about possible BSM phenomena can emerge starting from the observed discrepancies with respect to SM predictions \cite{CidVidal:2018eel}.
High precision experimental data, thanks to large luminosities and a great reduction of systematical errors, have been collected at the LHC and at the Tevatron so that the sensitivity to SM deviations depends crucially from the size of the theoretical uncertainties, which ought to be reduced.
\\
The inclusion of radiative corrections is mandatory in order to obtain accurate theoretical predictions for cross-sections and related kinematic distributions. Although the dominant contributions are due to strong interactions, the inclusion of electroweak effects become essential in the aim of reaching a percent or sub-percent level precision, given that $\alpha \sim \alpha_S^2$.
\\
In this paper, we consider the transverse-momentum ($q_T$) differential distributions of $W$ and $Z$ bosons, which are particularly relevant among the various kinematic distributions. Specifically, an accurate analysis of the $Z$ boson spectra gives us important information about the mechanism of $W$ boson production. Moreover, a detailed knowledge of the $q_T$ spectrum of the $W$ boson at small and intermediate values of $q_T$ is crucial for a precise measurement of the $W$ boson mass~\cite{Rottoli:2023xdc,CDF:2013dpa, ATLAS:2017rzl, LHCb:2021bjt, CDF:2022hxs}.
\\
Nonetheless, in the low transverse-momenta region ($q_T \ll m_V$), where the majority of events is produced, large logarithmic corrections of the type $\alpha_S^n \, \text{ln}^{m}(m_V^2/q_T^2)$, which are originated by   soft and/or collinear initial-state partonic radiation, ruins the convergence of the fixed-order perturbative expansion in $\alpha_S$. A systematic all-order evaluation and resummation of the logarithmic-enhanced terms must be performed in order to obtain reliable predictions.  
\\
In the last years, a consistent QCD resummation formalism has been developed in Refs.\,\cite{Bozzi:2003jy,Bozzi:2005wk,Catani:2013tia}, and nowadays $q_T$ differential distributions can be computed at high
perturbative accuracy with theoretical precision at the percent level \cite{Camarda:2019zyx,Camarda:2021ict}. At such level of
precision the QED and EW effects needs to be included.
In Ref. \cite{Cieri:2018sfk} a combined approach for QCD and QED $q_T$ resummations has been developed for $Z$ boson production showing percent level effects due to QED effects. 
\\
In this work, we extend the approach of Ref. \cite{Cieri:2018sfk}, which is valid for electrically neutral high mass systems, in order to consider electrically charged final states
and, in particular, we consider the case of $W$ boson production.
We include resummation
effects at next-to-leading accuracy (NLL) in QED and next-to-next-to-leading accuracy (NNLL) 
in QCD, also including, at small $q_T$, fixed-order corrections at one loop in the EW theory and at two loops in QCD.
Our results have been matched respectively with the $\mathcal{O}(\alpha^2)$ and $\mathcal{O}(\alpha_S^2)$ fixed-order results in QED and QCD at intermediate-large values of $q_T$. 
\\
In the case of $W$ boson production a direct abelianization of the QCD $q_T$ resummation
formalism for colourless systems is not possible because of the electromagnetic charge of the $W$ boson in the final state and the ensuing soft radiation. Therefore we considered the $q_T$ resummation formalism for heavy-quark
production in QCD \cite{Catani:2014qha} and we adapted it for the case of massive electromagnetic
charges. An additional complication which appears in the case of $W$ production is related to the
fact that next-to-leading fixed-order corrections in QED cannot be included in a trivial way without
breaking the gauge invariance of the calculation \cite{Wackeroth:1996hz}. We avoided this issue by  combining our resummed results with the full, gauge invariant, EW corrections at one loop. Therefore our results include, in the small $q_T$ region, the EW corrections at fixed-order and the effect of logarithmically enhanced QED radiation to all perturbative orders.
\\
The outline of this paper is the following. In Sec. \ref{sec:Introductionqt} we present a brief description of the $q_T$-resummation formalism of Refs. \cite{Catani:2013tia}. Then, we discuss the combination of QED and QCD corrections to the transverse-momentum resummation formalism in Sec. \ref{sec:CombinedQEDQCD}, recalling previous developments presented in Ref. \cite{Cieri:2018sfk}. After that, in Sec. \ref{sec:ResultsW}, we implement the explicit calculation for the case of $W$-production and describe the associated phenomenology for Tevatron and LHC in Sec. \ref{ssec:phenoresults}. In Sec. \ref{ssec:WZratio}, we compare the $q_T$ spectrum for $W$ and $Z$-boson production using the formalism described in Sec. \ref{sec:CombinedQEDQCD} and discuss the scale uncertainties propagation due to combined QED-QCD effects. Finally, we present the conclusions and depict possible paths for possible improvements in Sec. \ref{sec:Conclusions}.

\section{QCD transverse-momentum resummation}
\label{sec:Introductionqt}
In the context of high-energy physics, the calculation of cross-sections customary relies on the factorization theorem \cite{Collins:1989gx}. 
In this paper we consider the production of a vector boson $V$ ($V=W,Z,\gamma^*$), with invariant mass 
$M^2$, at hadron colliders. The associated vector boson transverse-momentum ($q_T$) differential cross-section describing a collision with hadronic centre-of-mass energy $\sqrt{s}$ is factorized as
\beqn
\nn \frac{d \sigma_{h_1 h_2 \to V}}{dq_T^2}(q_T,M,s) &=& \sum_{a_1, a_2} \int dx_1 dx_2 \, f_{a_1/h_1}(x_1,\mu_F^2) \,  f_{a_2/h_2}(x_2,\mu_F^2) \, 
\\ &\times& \frac{d \hat{\sigma}_{a_1 a_2 \to V}}{dq_T^2}(q_T,M,\hat{s};\muFsq) \, ,
\label{eq:Master1}
\eeqn
with $f_{a/h}(x,\mu_F^2)$ the parton distribution functions (PDFs) associated to the probability of extracting a parton of flavour $a$ from an hadron $h$ with longitudinal momentum fraction $x$, at the factorization scale $\muFsq$. 
The partonic centre-of-mass energy is given by $\hat{s}=\sqrt{x_1 x_2 s}$, under the assumption of massless colliding partons, and $d\hat{\sigma}$ denotes the partonic cross-section.

The partonic cross section $d\hat{\sigma}$ can be computed within perturbation theory, whereas the PDFs are extracted from data or modelled by non-perturbative methods. The dependence of the renormalization scale $\muRsq$ is included in the partonic cross-section, and it can be accounted for by a suitable redefinition of the running couplings.

Regarding the partonic cross-section, it is well known that fixed-order perturbation theory fails to accurately describe the low-$q_T$ region, $q_T\ll M$. This is because the presence of logarithmically enhanced terms, proportional to $\log(q_T^2/M^2)$ which grow faster than the suppression introduced by the higher-powers of the QCD coupling $\as$. For this reason, $q_T$ resummation
formalism have been developed in QCD in order to rearrange the perturbative series and resum it collecting to all orders the terms proportional to $\as^{n} \,  \log^{m}(q_T^2/M^2)$. 

The resummation of the perturbative expansion is achieved by explicitly splitting the partonic cross section as
\beq
\frac{d \hat{\sigma}}{dq_T^2} = \frac{d \hat{\sigma}^{\rm (res.)}}{dq_T^2} + \frac{d \hat{\sigma}^{\rm (fin.)}}{dq_T^2} \, ,
\label{eq:Division}
\eeq
with all the logarithmically enhanced contributions embodied within $d \hat{\sigma}^{\rm (res.)}$. More precisely, the finite component is defined in such a way that
\beq
\lim_{Q_T \to 0} \, \int^{Q_T^2}_{0} \, dq_T^2 \, \frac{d \hat{\sigma}^{\rm (fin.)}}{dq_T^2} = 0 \, .
\label{eq:DefFINITE}
\eeq
By performing the resummation in the impact parameter space $b$~\cite{Parisi:1979se}, conjugated to $q_T$, the resummed component can be written  as~\cite{Bozzi:2005wk}
\beqn
\frac{d\hat{\sigma}_{a_1 a_2 \to V}^{\rm (res.)}}{dq_T^2}(q_T,M,\hat{s};\muFsq) &=&  \frac{M^2}{\hat{s}} \, \int_0^\infty db \,\frac{b}{2} \,  J_0(b\, q_T)\, {\cal W}_{a_1 a_2}^V(b,M,\hat{s};\muFsq) \,,
\label{eq:Resumado1}
\eeqn
with $J_0$ the 0th-order Bessel function.
The function  ${\cal W}_{a_1a_2}^V$ can be organized in an exponential structure and, by considering the 
Mellin $N$-moments ${\cal W}^V_{a_1a_2,N}$
with respect to $z=M^2/\hat s$ at fixed $M$, in the simplified flavour-diagonal case ($a_1a_2=c\bar c$) it reads \cite{Bozzi:2005wk}: 
\beqn
\label{eqW}
      {\cal W}^V_{N}(b,M;\mu_F) =
      \hat{\sigma}^{(0)}_{V}(M)\,
      {\cal H}_{N}^V(\as;\mu_R^2,\mu_F^2,Q^2)
\times  \exp \left\{ {\cal G}_{N}(\as, L; \mu_R^2,Q^2) \right\} \,,
\label{eq:Resumado2}
\eeqn
where $\hat\sigma^{(0)}_{V}$ is the lowest-order partonic cross section of the hard-scattering process $h_1h_2\to V$. The hard-collinear function ${\cal H}_N^V$ includes contributions from the process-dependent hard-virtual corrections and can be expanded in powers of $\as = \as(\muRsq)$ as
\beqn
{\cal H}_{N}^V(\as) &=& 1+\sum_{n=1}^{+\infty}  \left(\frac{\as}{\pi}\right)^{n} {\cal H}_{N}^{V\,(n)} \, .
\label{eq:calHexpansion}
\eeqn
The exponent ${\cal G}_N$ resums  
in an universal (i.e. process-independent) way
all the logarithmically enhanced terms and can systematically expanded as:
\beqn 
{\cal G}_{N}(\as, L) 
&=&   -\int^{Q^2}_{b_0^2/b^2} \frac{d q^2}{q^2} \left(A(\as(q^2)) \log\left(\frac{M^2}{q^2}\right) + 
\widetilde{B}_{N}(\as(q^2)) \right)  \nonumber \\
&=& L \, g^{(1)}(\as L) \, + g_N^{(2)}(\as L)  
+ \sum_{n=3}^{+\infty} \left(\frac{\as}{\pi}\right)^{n-2} g_N^{(n)}(\as L)\,, 
\label{eq:calGexpansion2}
\eeqn 
where the functions $A(\as)$ and $\widetilde{B}_{N}(\as)$ are perturbative functions with a customary fixed-order expansion in powers 
of $\as$~\cite{Bozzi:2005wk}.
The logarithmic variable $L$ reads, in terms of the resummation scale $Q$ and $b_0=2 e^{-\gamma_E}$ ($\gamma_E = 0.5772 \cdots$ is the Euler number), 
\beq
L = \log\left(\frac{b^2 \, Q^2}{b_0^2}+1\right) \, ,
\label{eq:DefL}
\eeq
which corresponds, in the $b$-space at large $b$, to the logarithmically-enhanced contributions associated to the low-$q_T$ region. The scale $Q$~\cite{Bozzi:2005wk} is introduced with the purpose of parametrizing the arbitrariness in the factorized structure of Eq. (\ref{eq:Resumado2}). 

Truncating Eq. (\ref{eq:calGexpansion2}) up to the first term defines the leading-logarithm (LL) approximation, up to the second term defines the NLL approximation, then NNLL and so on.

In the case of transverse-momentum resummation for processes with colourful final states the explicit structure of the functions ${\cal H}_{N}$ and ${\cal G}_{N}$ differs with respect to the case of production of colourless final states.
The $q_T$ resummation formalism for colourful final states requires a colour space
diagonalization of the relevant soft-anomalous dimension which has been worked out, 
in the case of heavy-quark pair production up to NLL, in Ref.~\cite{Catani:2014qha}.
In particular for heavy-quark pair production at NLL accuracy the function
${\cal G}_{N}$ contains an additional component due to soft wide-angle radiation from the heavy quarks 
in the final state and from initial/final-state colour interference~\cite{Catani:2014qha}. Analogously due to colour correlations produced by soft-parton radiation, the hard-collinear function ${\cal H}_{N}$ is a colour space matrix. 

\section{Combined QED and QCD transverse-momentum resummation for\\\mbox{charged final states}}
\label{sec:CombinedQEDQCD}
In Ref.~\cite{Cieri:2018sfk} some of us extended the QCD resummation formalism for colourless final states in order to deal with the {\itshape combined} resummation of QED and QCD radiation in the case of colourless and electrically neutral final states. This combined resummation method has been obtained in two steps: \emph{(i)} by the corresponding \emph{abelianization}~\cite{deFlorian:2015ujt,deFlorian:2016gvk} of the QCD resummation formalism \cite{Bozzi:2005wk}; \emph{(ii)} by a consistent combination of the QED and QCD resummation effects. In particular in Ref. \cite{Cieri:2018sfk} it has been considered the explicit case of $Z$ boson production at hadron colliders up to NNLL in QCD and NLL in QED.
In this paper, we extend the formalism of Ref.~\cite{Cieri:2018sfk} in order to deal with 
the more general case of colourless but electrically charged final states. 

In order to generalise the combined QED and QCD $q_T$ resummation for the case of a charged final state, we need to take into account the effect of additional QED soft wide-angle radiation.  
We thus start from the QCD resummation formalism of Ref.~\cite{Catani:2014qha} developed for heavy-quark pair production and we adapt it to the case of QED resummation for high-mass charged systems, considering the particular case of $W$ boson hadroproduction.
To this end we need to consider the following modifications:  \emph{(i)} the replacement of the two particle final state kinematics (the quark-antiquark pair) by a single particle production kinematics;
\emph{(ii)} the abelianization of the QCD result, taking into account the absence of non-abelian color correlations produced by initial/final-state interference in the QED case.

Applying the abelianization procedure~\cite{deFlorian:2015ujt,deFlorian:2016gvk,Cieri:2018sfk} to Eqs. (15-18) of Ref.~\cite{Catani:2014qha} we obtain that the exponentiation
of large logarithmic corrections receive contributions from a QED soft radiation factor which can be written as:
\beqn 
\Delta(\alpha;Q,b) &=& \exp{\left\{-\int_{b_0^2/b^2}^{Q^2}\frac{dq^2}{q^2} D'(\alpha(q^2))\right\}},
\label{eq:Delta}
\eeqn
which is specific of charged high-mass system production and it is due to QED soft non-collinear (wide angle) radiation from the underlying subprocess (in our specific case, $q_f\bar{q}_{f'}\to W^\pm$). In particular, soft non-collinear radiation originates from final state emissions from the final
state charged system and from initial/final-state interferences. The soft factor in Eq. \eqref{eq:Delta} involves an integration over the 
transverse-momentum range $1/b\lesssim q_T \lesssim Q$ giving rise to additional enhanced logarithmic corrections of the type
$\alpha^n\ln{Qb}^k$ which are resummed to all orders in an exponential form. The function $D'(\alpha)$ has the following standard perturbative expansion in power of $\alpha$
\beqn
D'(\alpha) &=& \frac{\alpha}{\pi} \, D'^{(1)} + \sum_{n=2}^{+\infty} \, \left(\frac{\alpha}{\pi}\right)^n \, D'^{(n)} \,.
\label{eq:DWEXPANSION}
\eeqn

Therefore the abelianization of QCD factor $\mathcal{G}_N$ (previously introduced in Eq. \eqref{eq:calGexpansion2}) which takes into account corrections from QED emissions for the production of a charged high mass final state, is given by 
\beqn 
{\cal G}_{N}'(\alpha, L)
&=&   -\int^{Q^2}_{b_0^2/b^2} \frac{d q^2}{q^2} \left(A'(\alpha(q^2)) \log\left(\frac{M^2}{q^2}\right) + 
\widetilde{B}'_{N}(\alpha(q^2))+D'(\alpha(q^2)) \right)  \nonumber \\
&=& L \; g'^{(1)}(\alpha L)+ g_N'^{(2)}(\alpha L) 
+\sum_{n=3}^{+\infty} \left(\frac{\alpha}{\pi}\right)^{n-2} g_N'^{(n)}(\alpha L)\,,
\label{eq:Gprim}
\eeqn
where we note the presence of the previously introduced function $D'(\alpha)$, while the functions 
$A'(\alpha)$ and $\widetilde{B}'_{N}(\alpha)$ are related to QED radiation from initial state~\cite{Cieri:2018sfk} and can be expanded as:
\beqn
A'(\alpha) &=&  \frac{\alpha}{\pi} A'^{(1)}+ \left(\frac{\alpha}{\pi}\right)^2 A'^{(2)} + \sum_{n=3}^{+\infty} \left(\frac{\alpha}{\pi}\right)^n A'^{(n)} \, ,
\label{eq:AccprimeEXPANSION}
\\ \widetilde{B}'_{N}(\alpha) &=& \frac{\alpha}{\pi} \widetilde{B}'^{(1)}_{N}+ \sum_{n=2}^{+\infty}  \left(\frac{\alpha}{\pi}\right)^n \widetilde{B}'^{(n)}_{N} \,. 
\label{eq:ABEXPANSION}
\eeqn
As can be seen from Eq. (\ref{eq:Gprim}), the function $D'(\alpha)$ resums single-logarithmic corrections and it thus starts to contribute at NLL accuracy, similarly to the flavour-conserving 
collinear radiation function $B'_{N}(\alpha)$.
Therefore the structure of the exponential factor in Eq. (\ref{eq:Gprim}) which resums the large logarithmic corrections from QED radiation in the case of high-mass charged systems can be obtained from the case of neutral systems (see Eq. (2.7) in Ref. \cite{Cieri:2018sfk}) with the replacement:
\beqn
\widetilde{B}'_{N}(\alpha) \to \widetilde{B}'_{N}(\alpha)+D'(\alpha)\,.
\label{eq:BtoBD}
\eeqn
The presence of logarithmic effects from soft wide-angle emissions (through the function $D'(\alpha)$ in Eq. (\ref{eq:Gprim})) has also consequences in the determination of the {\itshape finite} component (see Eq. (\ref{eq:Division})), which is typically calculated from the fixed-order expansion of the {\itshape resummed} component. The substitution in Eq. (\ref{eq:BtoBD}) also holds in the case of the finite component.

The coefficient $D'^{(1)}$, which is not present in the case of the production of chargeless final states, can be obtained by a suitable abelianization of the soft anomalous dimension matrix (see Eqs. (15)-(17) in Ref. \cite{Catani:2014qha}). This resummation coefficient depends on the squared charge of the final state system $e_V^2$~\footnote{The electric charges are defined in units of $e$, where $-e<0$ is the electron
charge.} ($e_W^2=1$ in the case of $W$ production) and it reads
\begin{align}
D'^{(1)}=-\frac{e_V^2}{2}\,.
\label{eq:D1}
\end{align}

The resummation coefficients related to initial-state emissions, $A'^{(1)}$, $A'^{(2)}$ and $\widetilde{B}'^{(1)}_{N}$, have been obtained in Ref.~\cite{Cieri:2018sfk} (see Eqs. (2.19)-(2.20) of Ref.~\cite{Cieri:2018sfk}) from the corresponding coefficients in QCD~\cite{Kodaira:1981nh,Kodaira:1982az,Kodaira:1982cr,Catani:1988vd} for the case of the production of chargeless systems (e.g.\ for $Z$ boson production). In such case they are proportional to the square of the electric charge $e_q^2$ of
the initial state partons of the subprocess $q\bar{q}\to Z$. In the case of $W$ boson production the same coefficients can be obtained by replacing the squared charge  by the average of the squared charges of the initial state partons of the sub-process $q_f\bar{q}_{f'}\to W$:
\begin{align}
e_q^2 \to \frac{e_{q_f}^2+e_{\bar{q}_{f'}}^2}{2}=\frac{5}{18} \, .
\end{align}
The explicit values of the coefficients are:
\begin{align}
\label{eq:A1}
A'^{(1)} &= \frac{e_{q_f}^2+e_{\bar{q}_{f'}}^2}{2} \,,\\
\label{eq:A2}
A'^{(2)} &= -\frac{5}{9}\,\frac{e_{q_f}^2+e_{\bar{q}_{f'}}^2}{2} \,N^{(2)},\\
\label{eq:B1}
\widetilde B_{N}'^{(1)} &= B'^{(1)} + \gamma_{q_fq_f,N}'^{(1)}+ \gamma_{\bar{q}_{f'}\bar{q}_{f'},N}'^{(1)}\,,
\end{align}
with
\begin{align}
N^{(2)} &= 3 \sum_{q=1}^{n_f} e_q^2 + \sum_{l=1}^{n_l} e_l^2\,,\\
\label{eq:B1p}
  B'^{(1)} &= -\frac{3}{2} \,\frac{e_{q_f}^2+e_{\bar{q}_{f'}}^2}{2} \,,\\ 
\gamma_{qq,N}'^{(1)}&=e_q^2\,\left(\frac34+\frac{1}{2N(N+1)}-\gamma_E-\psi_0(N+1)\right)\,,\\
\gamma_{q\gamma,N}'^{(1)}&=\frac32\,e_q^2\,\frac{N^2+N+2}{N(N+1)(N+2)}\,,
\end{align}
where $\psi_0(N)$ is the digamma function and $\gamma_{ab,N}'^{(1)}$ are the leading-order (LO) anomalous dimensions in 
QED~\footnote{The anomalous dimension $\gamma_{q\gamma,N}'^{(1)}$ enters at the NLL in the general multiflavour case 
(see Appendix A of Ref.~\cite{Bozzi:2005wk}).},
$n_f$ ($n_l$) the number of quark (lepton) flavours and $e_q$ ($e_l$) the quark (lepton) 
electric charges ($e_q=2/3$ for up-type quarks, $e_q=-1/3$ for down-type quarks, $e_l=-1$ for leptons).

The knowledge of the resummation coefficient ${D}'^{(1)}$ in Eq. (\ref{eq:D1}), together with the coefficients
$A'^{(1)}$, $A'^{(2)}$ and $\widetilde{B}'^{(1)}$ in Eqs. (\ref{eq:A1}-\ref{eq:B1}), 
is sufficient to reach the full NLL accuracy for the resummed component in QED.

The results obtained for the resummation coefficients have been crosschecked in App.~\ref{App:RadiativeFSR} where we performed the expansion at small $q_T$ of the fixed-order $q_T$ distribution and we extracted the resummation coefficients confirming the results shown in this Section. 

We now consider the QED fixed-order contributions included in the hard-collinear function in Eq. (\ref{eq:calHexpansion}). 
We start considering 
the abelianization of the QCD infrared (IR) subtraction operator of Ref.~\cite{Catani:2014qha} and we obtain the following
QED IR subtraction operator:
\beqn
\widetilde{I}\,'_{V}(\epsilon,M^2) = \frac{\alpha(\mu_R)}{2\pi} \widetilde{I}\,'^{(1)}_{V}(\epsilon,M^2/\mu_R^2)
+\sum_{n=2}^{+\infty}\,\left(\frac{\alpha(\mu_R)}{2\pi}\right)^n \widetilde{I}\,'^{(n)}_{V}(\epsilon,M^2/\mu_R^2)\, 
\label{Iall}
\eeqn
with
\beqn
\widetilde{I}\,'^{(1)}_{V}(\epsilon,M^2/\mu_R^2)&=&-\left(\frac{M^2}{\mu_R^2}\right)^{-\epsilon}
\Bigg\{\left(\frac{1}{\epsilon^2}+\frac{i\pi}{\epsilon}-\frac{\pi^2}{12}\right)\frac{{e_{q_f}^2+e_{\bar{q}_{f'}}^2}}2+\frac{\gamma_{q_f}'+\gamma_{\bar{q}_{f'}}'}{2\epsilon}\nn\\
&&+\frac{e_V^2}{2\epsilon}(1-i\pi)\Bigg\} \, ,
\label{I1}
\eeqn
where the coefficient $\gamma_q'=3 e_q^2/2$ originates from hard-collinear initial-state radiation
while the last term proportional to $e_V^2$ originates from soft wide-angle radiation from the final state charged system.
Following the QCD case~\cite{Catani:2013tia,Catani:2014qha}, the subtraction operator in Eq. (\ref{I1}) allows us to define, starting from the renormalized IR divergent all-loop amplitude $\mathcal M_{V}$, an IR finite hard-virtual amplitude 
\beqn
\widetilde{\mathcal M}_{V}= (1-\widetilde{I}\,'_{V}(\epsilon,M^2))\mathcal M_{V}\, .
\label{Mtilde}
\eeqn
At one-loop Eq. (\ref{Mtilde}) reads
\beqn
\widetilde{\mathcal M}_{V}^{(1)}= \mathcal M_{V}^{(1)} -\widetilde{I}\,'^{(1)}_{V}(\epsilon,M^2/\mu_R^2) \mathcal M_{V}^{(0)}\,,
\label{Mtilde1}
\eeqn
where ${\mathcal M}^{(0)}_{V}$ and ${\mathcal M}^{(1)}_{V}$ are respectively the lowest-order and the one-loop parton-level scattering amplitude for the scattering process $h_1h_2\to V$.
In turn the knowledge of the hard-virtual amplitude $\widetilde{\mathcal M}_{V}$ is sufficient to determine the process-dependent hard-virtual coefficient~\cite{Catani:2013tia}:
\beqn
H'^{V}(\alpha)=1+ \sum_{n=1}^{+\infty} \left(\frac{\alpha}{\pi}\right)^n {H'}^{V\,(n)}= \frac{|\widetilde{\mathcal M}_{V}|^2}{|{\mathcal M}^{(0)}_{V}|^2}\,,
\label{Hstraight}
\eeqn
which encodes the process-dependent part of the hard-collinear coefficient $\mathcal{H}_N'^{V}$ Eq. (\ref{eq:calHexpansion}) in QED.

As already mentioned,
in the case of $W$ production, the one-loop corrections in QED cannot be included in a trivial way without
breaking the gauge invariance of the results~\cite{Wackeroth:1996hz}. This issue is relevant only for the fixed-order corrections
and it does not affect the all-order resummation of enhanced QED logarithmic effects. We solved this issue by including in our results the full EW corrections at one loop in the scattering amplitude ${\mathcal M}^{(1)}_{V}$. To be consistent, we included the one-loop EW corrections also in the case of $Z$ boson production, even if in this case the pure QED corrections can be defined in a straightforward way.

The explicit results for the (not vanishing) NLO hard-collinear functions 
${\cal H}_{a_1 a_2, N}'^{V\,{(1)}}$ we have included in our calculation are:
\begin{align}
\label{h1qed}
{\cal H}_{q_{f}\bar{q}_{f'} \leftarrow q_{f}\bar{q}_{f'},N}'^{V \,(1)}&=\frac{e_{q_{f}}^2+e_{\bar{q}_{f'}}^2}{2} \, \left(\frac{1}{N(N+1)} +H'^{V\,(1)}  \right) \,,\\
{\cal H}_{q_{f}\bar{q}_{f'} \leftarrow \gamma \bar{q}_{f'},N}'^{V \,(1)}&=\frac{3\, e_{q_{f}}^2}{(N+1)(N+2)} \,,\\
{\cal H}_{q_{f}\bar{q}_{f'} \leftarrow \bar{q}_{f}\gamma ,N}'^{V \,(1)}&=\frac{3\, e_{\bar{q}_{f'}}^2}{(N+1)(N+2)} \,,
\end{align}
where the coefficient $H'^{V\,(1)}$ for $V=\gamma^*/Z,W$ has been obtained through Eqs.~(\ref{Mtilde}-\ref{Hstraight})
from the knowledge of the EW one-loop amplitudes for the processes
$q\bar q \to Z$ and $q_f{\bar q}_{f'} \to W$~\cite{Behring:2020cqi,Bonciani:2021iis}.

Finally we performed the matching at large $q_T$ by evaluating the finite part of the partonic cross section in Eq.~(\ref{eq:Division}) 
starting from the computation of the partonic cross section for $V+\gamma$ at leading-order in QED 
and subtracting from it the perturbative truncation of the resummed component at the same order:
\beq
\frac{d \hat{\sigma}^{\rm (fin.)}}{dq_T^2} = \Bigg[\frac{d \hat{\sigma}}{dq_T^2}\Bigg]_{\rm (f.o.)} -  
\Bigg[ \frac{d \hat{\sigma}^{\rm (res.)}}{dq_T^2}\Bigg]_{\rm (f.o.)}\,.
\label{eq:Division2}
\eeq

Having obtained the $q_T$ resummation formalism in QED for final state charged high-mass system we are able to
consider the combined QED and QCD resummation using the formalism developed in Ref.\,\cite{Cieri:2018sfk}
that we summarize below.

We combine the QED and QCD resummation formalism by
replacing the functions ${\cal W}_{N}^V$ and ${\cal G}_{N}$ in Eq.~(\ref{eq:Resumado2}) by their generalised expressions 
which include combined QCD and QED
effects through a double perturbative expansion in powers of $\as$ and of the electromagnetic coupling evaluated at the renormalization scale 
$\alpha=\alpha(\mu_R')$:
\begin{align}
\label{eqG2}
{\cal G}_{N}'(\as,\alpha, L)&={\cal G}_{N}(\as, L)
+
L \; g'^{(1)}(\alpha L)+ g_N'^{(2)}(\alpha L) 
+\sum_{n=3}^{+\infty} \left(\frac{\alpha}{\pi}\right)^{n-2} g_N'^{(n)}(\alpha L)\nn\\
&
+
g'^{(1,1)}(\as L,\alpha L)
+
\sum_{n,m=1 \atop n+m\neq 2}^{+\infty} \left(\frac{\as}{\pi}\right)^{n-1}\left(\frac{\alpha}{\pi}\right)^{m-1} g_N'^{(n,m)}(\as L,\alpha L)\;\;,
\end{align}
and
\begin{align}
\label{eqH2}
{\cal H}_N'^{V}(\as,\alpha)&= {\cal H}_N^{V}(\as)
+ \frac{\alpha}{\pi} \,{\cal H}_N'^{V \,(1)}
+\sum_{n=2}^{+\infty} \left(\frac{\alpha}{\pi}\right)^{n}\,{\cal H}_N'^{V \,(n)} \nn\\ 
&+\sum_{n,m=1}^{+\infty} \left(\frac{\alpha_S}{\pi}\right)^{n}\left(\frac{\alpha}{\pi}\right)^{m}\,{\cal H}_N'^{V \,(n,m)}\,.
\end{align}
The functional form of the functions $L\, g'^{(1)}$, $g_N'^{(2)}$ and $g'^{(1,1)}(\as L,\alpha L)$ 
can be found in Ref.~\cite{Cieri:2018sfk}.
We recall that the function $L\, g'^{(1)}$ resums to all order 
the LL contributions in QED, the function $g_N'^{(2)}$ the NLL ones and so on, while the terms $g'^{(1,1)}(\as L,\alpha L)$ and $g_N'^{(n,m)}(\as L,\alpha L)$ include respectively the leading and subleading mixed QCD-QED corrections.
In the case of the production of a charged high-mass system, the function $g_N'^{(2)}$ receive a contribution from
soft wide-angle QED radiation which has been included through the replacement in Eq. (\ref{eq:BtoBD}).
The coefficients ${\cal H}_N'^{V \,(n)}$ control the pure QED corrections while the coefficients ${\cal H}_N'^{V \,(n,m)}$
contains the mixed QCD-QED ones. 

Finally we have considered, in the factorization formula Eq. (\ref{eq:Master1}),
the inclusion of the photon parton density $f_{\gamma/h}(x,\mu_F^2)$ and the QED effects in the evolution of parton densities.

\section{Numerical results for W and Z boson production at hadron colliders}
\label{sec:ResultsW}
In this section, we present selected phenomenological predictions for $W$ and $Z$ boson $q_T$ distributions at the Tevatron and at the LHC. Special emphasis is given to the similarities and differences between charged and neutral weak boson production.

The resummation formalism, together with a consistent matching-procedure to describe a wide $q_T$ region with numerical stability and uniform accuracy, is encoded in the \textsc{Fortran} numerical program \texttt{DYqT} \cite{Cieri:2018sfk,Bozzi:2008bb,Bozzi:2010xn}. In particular, we fully include QED radiation at NLL accuracy, matched with fixed order results at NLO in EW theory combined with the QCD corrections at NNLL+NNLO accuracy.  

We use the following values for the electroweak input parameters~\cite{Workman:2022}:
\beq
\alpha(m_Z^2)=1/127.95 \, , \, \, m_W = 80.377 \, \text{GeV} \, , \, \, m_Z = 91.1876\, \text{GeV} \, ,
\eeq
and:
\begin{equation*}
|V_{CKM}| = \begin{pmatrix}
|V_{ud}| &   \, |V_{us}| &  |V_{ub}| \\
|V_{cd}| &   \, |V_{cs}| &  |V_{cb}| \\
|V_{td}| &   \, |V_{ts}| &  |V_{tb}| 
\end{pmatrix} = 
\begin{pmatrix}
0.97435 &   \, 0.22500 &  0.00369 \\
0.22486 &   \, 0.97349 &  0.04182 \\
0.00857 &   \, 0.04110 &  0.999118 
\end{pmatrix}.
\end{equation*}
We use the \textsc{NNPDF4.0} PDFs set at NNLO in QCD \cite{NNPDF:2021njg}, which includes the parton density of the photon and the LO QED in the PDFs evolution, as implemented in the LHAPDF framework \cite{Buckley:2014ana}. The strong coupling is evaluated at 3 loops with $\as(m_Z^2) = 0.118$ within the $\overline{MS}$ renormalization scheme. 
 
We work with $n_f = 5$ quarks flavours and $n_l = 3$ charged leptons in the massless approximation. The EW corrections depend also on the Higgs boson and top-quark masses and we use $m_H = 125.25 \,\, \text{GeV}$, $m_t = 172.5 \,\, \text{GeV}$, respectively.

Numerical predictions are complemented with a study of the associated perturbative uncertainty through a customary scale variation method.  
Since we are primarily interested in estimating the effects of yet unknown QED corrections, we perform variations of resummation ($Q'$) and renormalization ($\mu_R'$) QED scales, in the range $m_V/2 \leq \{\mu_R', 2Q'\} \leq 2 m_V$ with the constraint $1/2 \leq \{\mu_R'/Q'\} \leq 2$,
keeping the QCD scales fixed at the central values $\mu_F = \mu_R = 2 Q = m_V$.
In principle we could also vary the factorization scale related to the QED emissions ($\mu_F'$). However since the QCD and QED
factorization scales are fixed to be equal inside the evolution of parton densities \cite{NNPDF:2021njg}, we also fix 
$\mu_F' = \mu_F = m_V$~\footnote{An exhaustive scale-dependence study of QCD results up to NNLL+NNLO was carried out in Refs.~\cite{Bozzi:2008bb,Bozzi:2010xn,Catani:2015vma}.}.

\subsection{Phenomenological predictions}
\label{ssec:phenoresults}
We show now numerical results for $q_T$ distributions in charged and neutral weak-boson production at the Tevatron ($\sqrt{s} = 1.96 \, \text{TeV}$) 
and at the LHC ($\sqrt{s} = 13.6 \, \text{TeV}$). We considered predictions at low and intermediate transverse momentum region, where the resummation component is relevant and the bulk of the cross sections lies ($q_T\lesssim  40$\,GeV).

\begin{figure}[ht]
    \centering
    \includegraphics{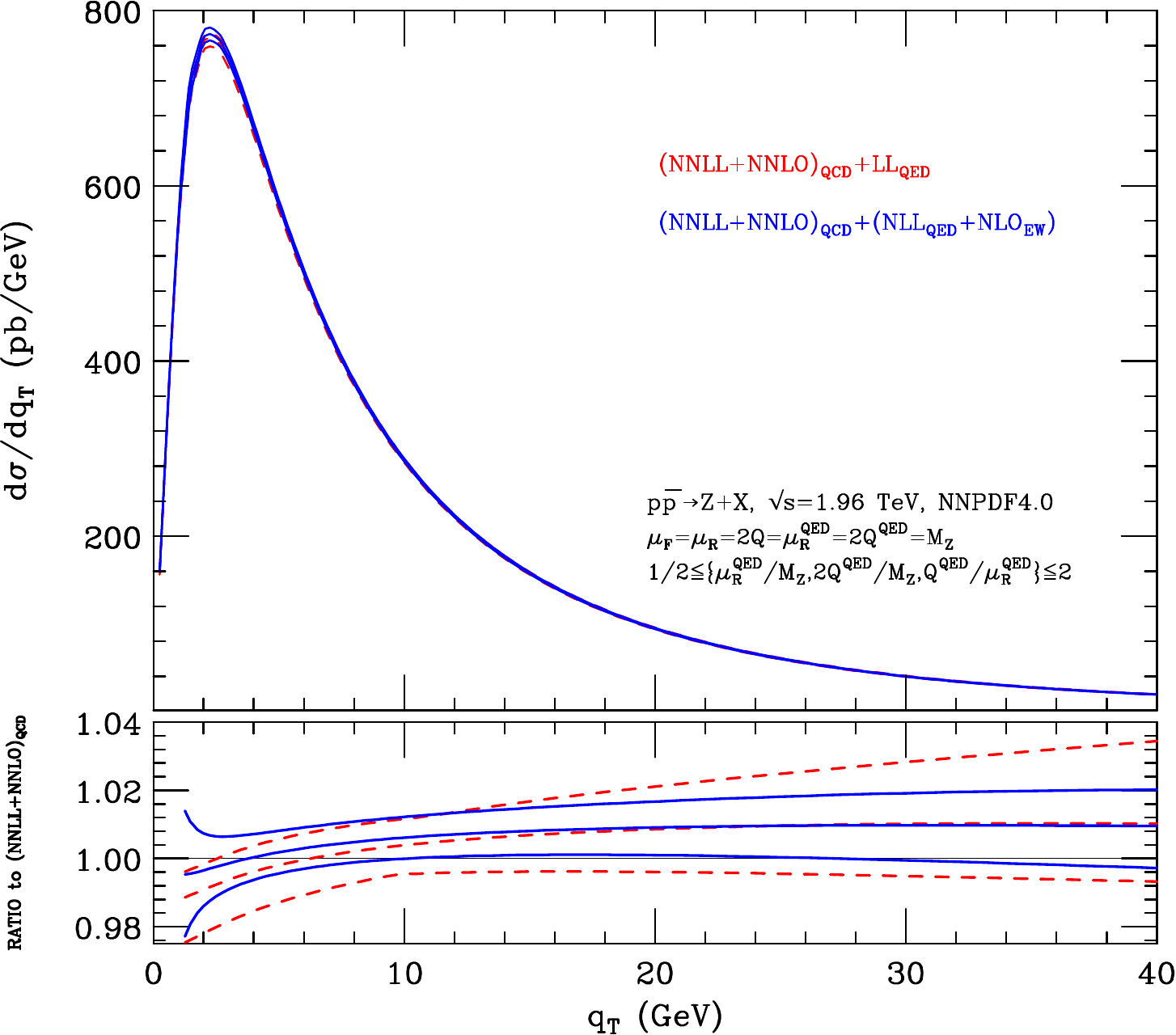}
    \caption{
    The $q_T$ distribution of $Z$ bosons produced at the Tevatron ($\sqrt{s}= 1.96$ TeV). The NNLL+NNLO results in QCD
    are combined with LL (red dashed) and NLL+NLO (blue solid) QED effects (with the inclusion of one-loop EW effects). 
    The uncertainty bands are obtained by performing the variation of the $\mu_R'$ and $Q'$ scales around their central value as described in the text. The lower panel shows the ratio of the results with respect to the standard NNLL+NNLO QCD result at central value of the scales.}
\label{ZqT-QCD+QED_tev_new}
\end{figure}

We start considering the (on-shell) $Z$ production in $p \bar{p}$ collisions at Tevatron energies
($\sqrt{s}= 1.96$ TeV). In Fig.~\ref{ZqT-QCD+QED_tev_new} we present the NNLL+NNLO QCD results combined with the
LL (red dashed) and NLL+NLO (blue solid) QED resummation which include the NLO EW corrections. 
The lower panel presents the ratio of our predictions with respect to the standard 
NNLL+NNLO QCD result at the corresponding
central scale.
As already observed in Ref.~\cite{Cieri:2018sfk}, we note that the resummation of the QED contributions at LL accuracy 
has the effect to make the $q_T$ spectrum slightly harder.  
The impact of the LL QED effects reaches the level of $\mathcal{O}(1\%)$. Thanks to the unitary constraint of the resummation formalism, the LL QED effects give vanishing contribution to the total cross
section affecting only the shape of the distribution by shifting part of the cross section to higher values of $q_T$. This physical effect is not unexpected and it is generated by soft and collinear QED emissions to all
orders. The NLL+NLO effects, for central values of the scales, are instead of $\mathcal{O}(0.5\%)$ level and are mainly concentrated in the the low $q_T$ region ($q_T\lesssim 10$\,GeV).
By considering the scale variation band, we observe that the LL QED effects have an uncertainty of around $2\%$ in the small $q_T$ region ($q_T\lesssim  10$\,GeV) which increases up to $4\%$ 
in the intermediate region $30\lesssim q_T\lesssim  40$\,GeV. 
The scale variation band is reduced by roughly a factor 2
with the inclusion of the NLL+NLO corrections. 

\begin{figure} [ht]
\label{ZqT-QCD+QED_lhc_new}
\begin{center}
\includegraphics[width = 150 mm, height = 120 mm ]{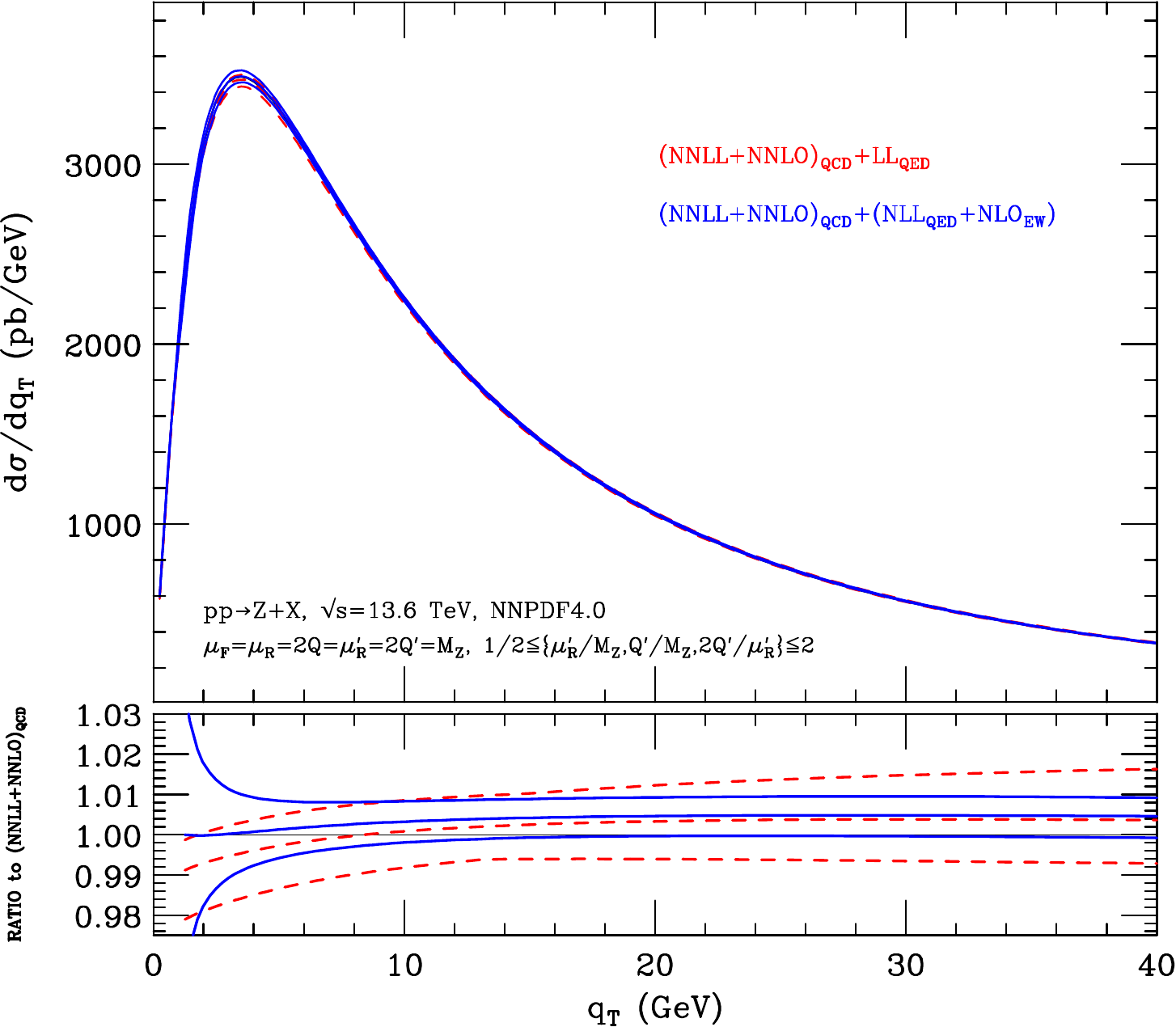}
\caption{The $q_T$ distribution of $Z$ bosons produced at the LHC ($\sqrt{s}= 13.6$\, TeV). The NNLL+NNLO results in QCD
    are combined with LL (red dashed) and NLL+NLO (blue solid) QED effects (with the inclusion of one-loop EW effects). 
    The uncertainty bands are obtained by performing the variation of the $\mu_R'$ and $Q'$ scales around their central value as described in the text. The lower panel shows the ratio of the results with respect to the standard NNLL+NNLO QCD result at central value of the scales.}
\end{center}
\end{figure}
In Fig.~\ref{ZqT-QCD+QED_lhc_new} we show the results for the $Z$ boson $q_T$ distribution at the LHC ($pp$ collisions at $\sqrt{s}= 13.6$\, TeV).
We observe that the effects of the QED contributions at the LHC are
qualitatively similar but slightly smaller with respect to the case of the Tevatron.
This lower sensitivity to QED contributions with respect to the QCD ones at the LHC is expected because of 
the greater available center-of-mass energy and the ensuing enhancement of the gluon luminosities with respect to the quark ones. 
The QED effects at LL accuracy  has the effect to make the $q_T$ spectrum harder giving a (negative)
$\mathcal{O}(1\%)$ contribution at small $q_T$ ($q_T\lesssim  5$\,GeV) and a (positive) $\mathcal{O}(0.5\%)$ contribution at $q_T\gtrsim  10$\,GeV.
The NLL effects are positive and at the level of $\mathcal{O}(0.5\%)$ (or below) for the entire $q_T$ region we have considered 
($q_T\lesssim  40$\,GeV). Concerning the perturbative uncertaintes, we observe that the LL uncertainty is around $2\%$ and the inclusion of the NLL+NLO corrections reduces the scale variation band by roughly a
factor 1.5-2. 

In both cases, at the Tevatron and the LHC, QED uncertainty is dominated by the renormalization scale at LL accuracy and resummation scale at NLL+NLO.

The results presented for $Z$ boson production are similar to those presented in Ref.~\cite{Cieri:2018sfk}. 
The calculation presented in this paper differs from the one in Ref.~\cite{Cieri:2018sfk} because we have included the
one-loop EW corrections in the hard factor and we used a different set of PDFs. The effect of EW loop corrections is extremely small (per-mille level effect)
and has been included for theoretical consistency\footnote{ 
We also note that the in Ref.~\cite{Cieri:2018sfk} a different version of the code \texttt{DYqT} for the NNLL+NNLO QCD results
have been used.}.

\begin{figure} [ht]
    \centering
    \includegraphics{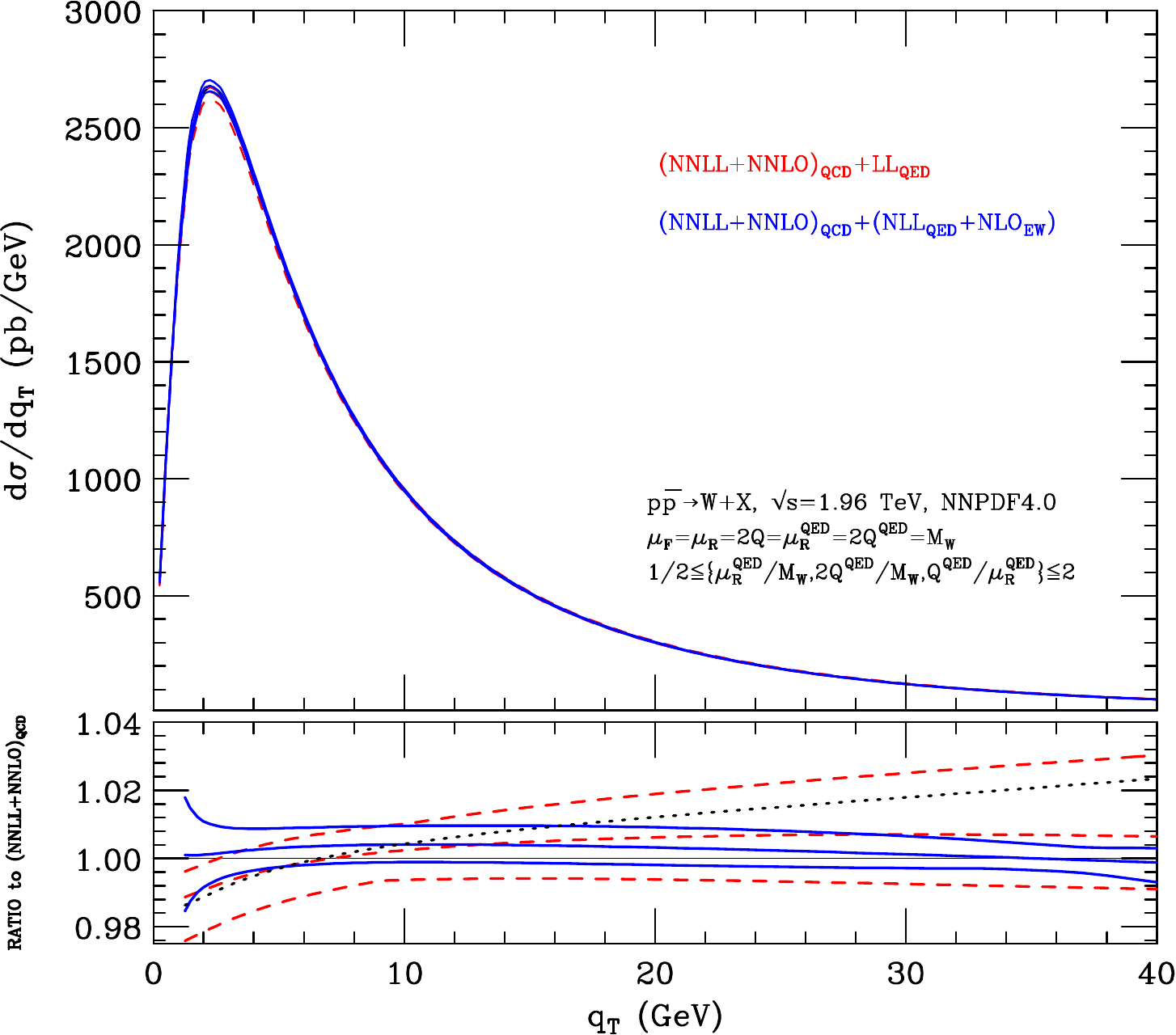}
    \caption{The $q_T$ distribution of $W$ bosons produced at the Tevatron ($\sqrt{s}= 1.96$ TeV). The NNLL+NNLO results in QCD
    are combined with LL (red dashed) and NLL+NLO (blue solid) QED effects (with the inclusion of one-loop EW effects). 
    The uncertainty bands are obtained by performing the variation of the $\mu_R'$ and $Q'$ scales around their central value as described in the text. The lower panel shows the ratio of the results with respect to the standard NNLL+NNLO QCD result at central value of the scales. The black dotted line shows NLL+NLO results obtained of removing the contribution of soft wide-angle radiation (i.e.\ setting $D_1'=0$) in the resummed component.}
    \label{WqT-QCD+QED_tev_new}
\end{figure}
In Fig.~\ref{WqT-QCD+QED_tev_new} we consider the novel predictions for $q_T$ distributions of $W$ bosons produced at the Tevatron, $\sqrt{s} = 1.96 \, \text{TeV}$ and at the LHC $\sqrt{s} = 13.6 \, \text{TeV}$. Since we are mainly interested on the higher-order QED effects we do not distinguish between $W^+$ and $W^-$ production and we consider both cases together. 
At LL accuracy, QED effects are similar to those discussed for the $Z$ boson production in Fig.~\ref{ZqT-QCD+QED_tev_new}, being driven by QED radiation from the initial state quarks.
The effect of NLL+NLO corrections in the case of $W$ boson production are instead different from the $Z$ boson case, giving
a $\mathcal{O}(1\%)$ positive (negative) correction for $q_T\lesssim  10$\,GeV ($q_T \gtrsim  20$\,GeV).
We recall that for $W$ boson production at the NLL there is the additional effect of soft wide-angle QED radiation
from the $W$ boson in the final state. 
In order to quantify the impact of such effect, we have considered the NLL+NLO prediction in which we removed
the contribution of the soft wide-angle radiation in the resummed component of Eq.(\ref{eq:Division}) 
(i.e.\ we set $D_1'=0$)\footnote{We observe, however, that we did not change the finite component  
where the effect of the $D_1'$ coefficient is necessary in order to cancel the divergent behaviour of the fixed-order term
for $q_T\to 0$ (see Eq. (\ref{eq:Division2})).}.

The NLL+NLO prediction without the effect of soft wide-angle QED radiation is shown in the lower panel of 
Fig.~\ref{WqT-QCD+QED_tev_new} (black dotted curve). We can see that soft wide-angle radiation has the effect to 
make the spectrum softer giving a positive 
$\mathcal{O}(1\%)$ contribution at small $q_T$ ($q_T\lesssim  5$\,GeV) and a negative $\mathcal{O}(1-2\%)$ contribution at $q_T\gtrsim  15$\,GeV. This effect can be expected by the fact that the coefficient
$D_1'$ is negative as the coefficient $B_1'$ and thus compensate the effect of the positive coefficients $A_1'$ and $A_2'$.
We note that the same qualitative effect due to the resummation of soft wide-angle radiation has been observed in Ref.~\cite{Catani:2018mei} in the case of QCD $q_T$ resummation for $t \bar{t}$ production. Concerning the perturbative uncertainties, we observe that the LL uncertainty is around $2-3\%$ and the inclusion of the NLL+NLO corrections reduces the scale variation band by roughly a factor 1.5-2 for $q_T\lesssim  20$\,GeV and up to a factor 3 for $q_T\gtrsim  30$\,GeV. 

 \begin{figure} [htbp]
    \centering
    \includegraphics{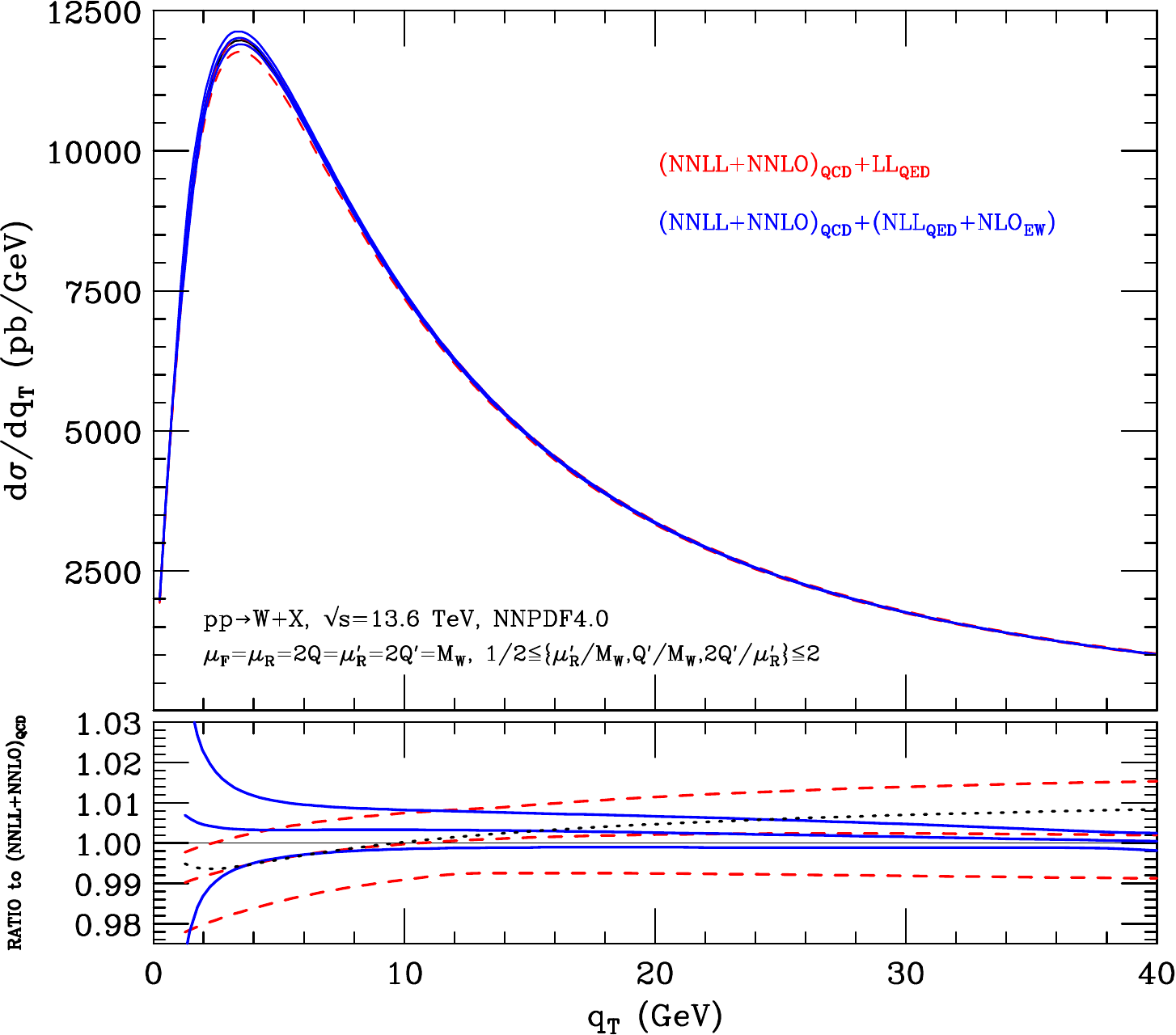}
    \caption{ The $q_T$ distribution of $Z$ boson produced at the LHC ($\sqrt{s}= 13.6$\, TeV). The NNLL+NNLO results in QCD
    are combined with LL (red dashed) and NLL+NLO (blue solid) QED effects (with the inclusion of one-loop EW effects). 
    The uncertainty bands are obtained by performing the variation of the $\mu_R'$ and $Q'$ scales around their central value as described in the text. The lower panel shows the ratio of the results with respect to the standard NNLL+NNLO QCD result at central value of the scales. The black dotted line shows NLL+NLO results obtained of removing the contribution of soft wide-angle radiation (i.e.\ setting $D_1'=0$) in the resummed component.}
    \label{WqT-QCD+QED_lhc13.6_new}
\end{figure}

Finally in Fig. \ref{WqT-QCD+QED_lhc13.6_new} we consider the predictions for $q_T$ distributions of $W$ bosons produced at the LHC, $\sqrt{s} = 13.6 \, \text{TeV}$. Also in this case the effect of LL QED resummation is to make the spectrum harder while the NLL+NLO effects goes in the opposite direction. The effect due to the resummation of soft wide-angle radiation is not negligible: the prediction obtained by setting $D_1'=0$ in the resummed component (black dotted line in the ratio panel of Fig. \ref{WqT-QCD+QED_lhc13.6_new}) decreases (increases) the cross section by $\mathcal{O}(1\%)$ at  $q_T\lesssim  5$\,GeV  ($q_T\gtrsim  20$\,GeV).
The NLL+NLO corrections reduce the LL scale variation band by a factor of 1.5-2 for $q_T\lesssim  20$\,GeV and up to a factor 4 for $q_T\gtrsim  30$\,GeV. 
The NLL+NLO uncertainty being of $\mathcal{O}(1.5\%)$ around the peak ($q_T\sim 3$\,GeV)
and decrease to about $\mathcal{O}(0.5\%)$ for $q_T\gtrsim 30$\,GeV.

As a general comment, we note that all the predictions in Figs.~(\ref{ZqT-QCD+QED_tev_new}-\ref{WqT-QCD+QED_lhc13.6_new}) 
show a good overlap of the LL and NLL+NLO scale uncertainty band, thus signalling a good behaviour
of the QED perturbative series. We also observe that the NLL+NLO bands tend to increase in the very small $q_T$ region 
($q_T\lesssim  2$\,GeV) where, however, we expect a sizeable role of truly non perturbative (NP) effects. Being mainly interested on (all-order) QED effects, in our results we did not introduce an explicit model for NP QCD contributions. In particular we used the so called  \emph{minimal prescription} \cite{Catani:1996yz,Laenen:2000de} in order to regularize the singularity of the resummed form factor in Eq.~(\ref{eq:calGexpansion2}) which occurs at large values of the impact parameter $b \sim 1/\Lambda_{QCD}$, where $\Lambda_{QCD}$ is the scale of the Landau pole of the perturbative QCD coupling.
 
\subsection{The ratio of $W$ and $Z$ transverse-momentum distributions}
\label{ssec:WZratio}
The measurement of the $W$ mass is directly affected by the uncertainty in the shape of the $W$ boson $q_T$ spectrum
which however is not directly experimental accessible due to neutrino in final state in the leptonic $W$ decay.  
Conversely, the $q_T$ spectrum of the $Z$ boson can be measured with great precision.
Therefore a precise theoretical prediction of the ratio of $W$ and $Z$ $q_T$ distributions, together with the measurement of the 
$Z$ boson $q_T$ spectrum, gives stringent information on the $W$ spectrum. We thus define the quantity
\begin{equation}
    R(q_T) = \frac{{\frac{1}{\sigma_{W}}}{\frac{d \sigma_{W}}{d q_T}}}{{\frac{1}{\sigma_{Z}}}{\frac{d \sigma_{Z}}{d q_T}} }.
\label{eq:RqT}
\end{equation}
The benefit of considering the prediction of the ratio of distributions, instead of the single one, lies also on the possible
reduction of the theoretical error, consequent to simplification of common (correlated) uncertainties. 

\begin{figure}
    \centering
    \includegraphics{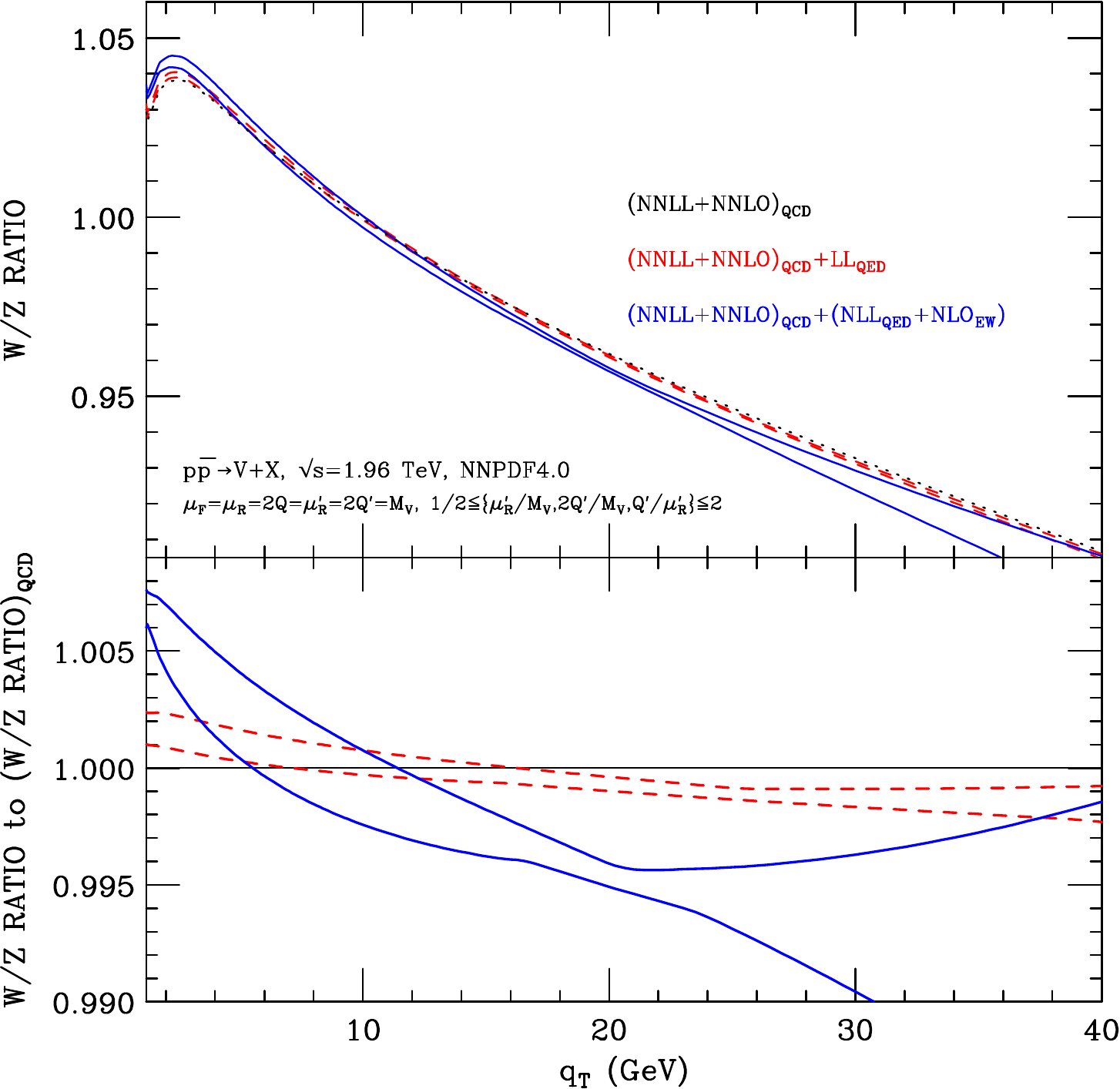}
    \caption{Normalized ratio of $W$ over $Z$ $q_T$-distribution in proton-antiproton collisions at Tevatron energy $\sqrt{s} = 1.96 \, \text{TeV}$. The NNLL+NNLO results in QCD are combined with LL (red dashed) and NLL+NLO (blue solid) QED effects (with the inclusion of one-loop EW effects). The uncertainty bands are obtained by performing the variation of the $\mu_R'$ and $Q'$ scales around their central value in a correlated way as described in the text. The lower panel shows the ratio of the results with respect to the standard NNLL+NNLO QCD result at central value of the scales.}
    \label{WZqT-QCD+QED_tev_new}
\end{figure}

\begin{figure}
    \centering
    \includegraphics{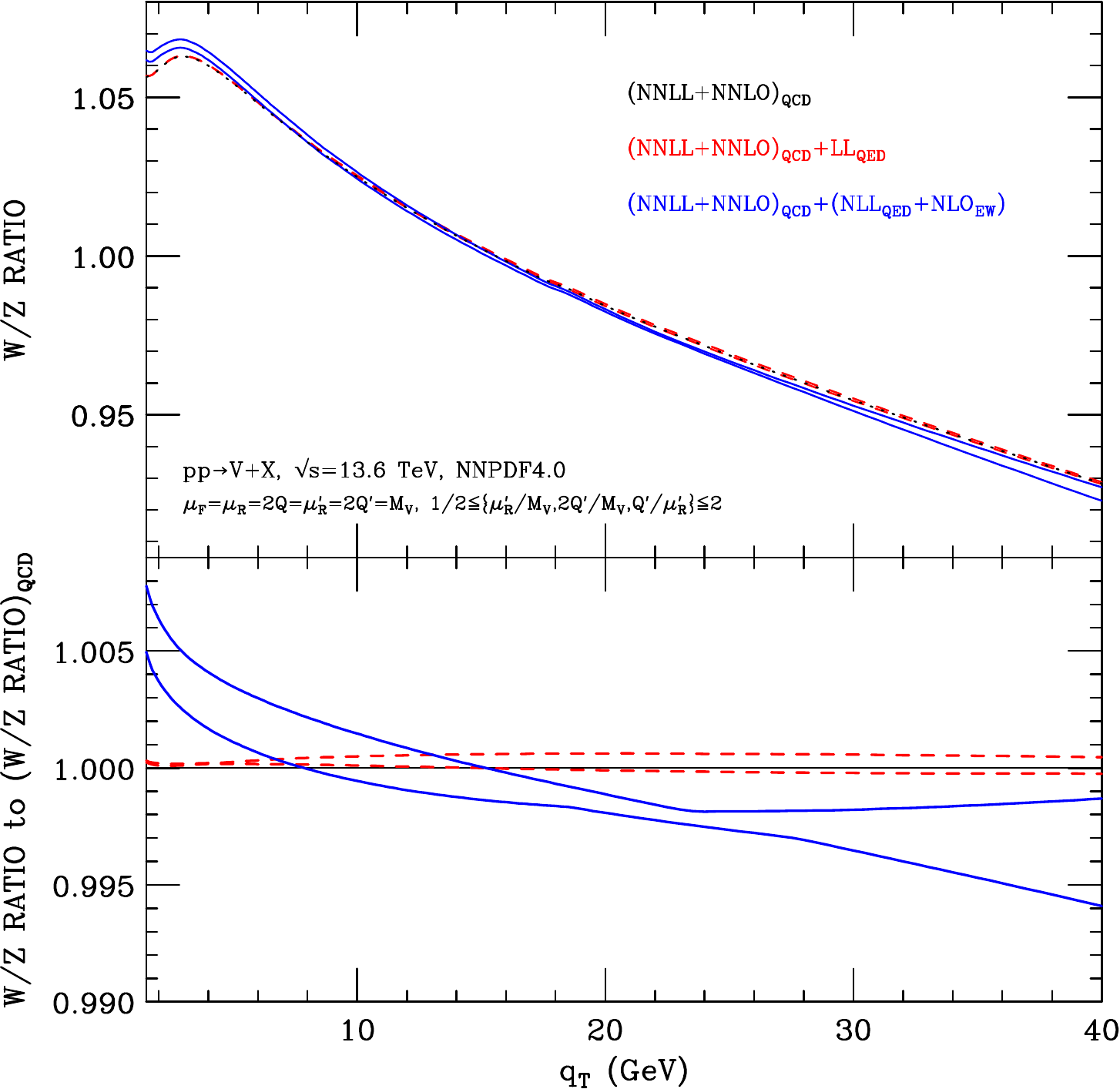}
    \caption{Normalized ratio of $W$ over $Z$ $q_T$-distribution in proton-proton collisions at LHC energy $\sqrt{s} = 13.6 \, \text{TeV}$. The NNLL+NNLO results in QCD are combined with LL (red dashed) and NLL+NLO (blue solid) QED effects (with the inclusion of one-loop EW effects). The uncertainty bands are obtained by performing the variation of the $\mu_R'$ and $Q'$ scales around their central value in a correlated way as described in the text. The lower panel shows the ratio of the results with respect to the standard NNLL+NNLO QCD result at central value of the scales.}
    \label{WZqT-QCD+QED_lhc13.6_new}
\end{figure}

In this section, we thus analyze the impact of QED corrections to the $R (q_T)$ distribution defined in Eq. \ref{eq:RqT}, showing phenomenological predictions for Tevatron ($\sqrt{s} = 1.96\, \text{TeV}$) in Fig.~\ref{WZqT-QCD+QED_tev_new}, as well as for LHC ($\sqrt{s} = 13.6\, \text{TeV}$) in Fig.~\ref{WZqT-QCD+QED_lhc13.6_new}. In the upper panels of the figures, the $R(q_T)$ distribution is shown, while in the lower panels we present the ratio of the quantity $R(q_T)$ with respect to the same quantity predicted in QCD, in order to asses the impact of QED radiation, as done in the previous section. As for the scale uncertainty band, we perform variations of $Q'$ and $\mu_R'$ QED scales as in the single distributions in a correlated way: that is, the scale combination of the $q_T$-distribution for the $W$ is the same of the one for the $Z$.

We start by analyzing the Tevatron case shown in Fig. \ref{WZqT-QCD+QED_tev_new}. From the lower panel, we can see that the LL QED contributions are at per-mille level, while scale variation band is of $\mathcal{O}(0.1 \%)$. The QED radiation makes the $R(q_T)$ distribution slightly
(up to few per-mille) softer: it
raises the QCD prediction for $q_T$ values smaller than $\sim  15 \, \text{GeV}$, lowering it for larger values of $q_T$. Overall the impact of LL QED corrections is strongly reduced with
respect to the single boson spectra. This reduction is the consequence of the similarity of LL QED emissions from the initial state quarks in $W$ and $Z$ production. 
Conversely at NLL, the impact of QED corrections is not suppressed, and that is reflected also by the size of the scale uncertainty band. The impact of NLL+NLO QED corrections is to make the
distribution softer at $\mathcal{O}(0.5-1\%)$ level. This is the combined effect of the $W$ distribution slightly softer and the $Z$ distribution harder. The scale variation band ranges from a minimum of $ \sim 0.1 \, \%$ at $q_T \sim 20 \, \text{GeV}$ to $1 \, \%$ level for $q_T \sim 35\, \text{GeV}$. 
The non-cancellation of common uncertainties can be explained by the  soft wide-angle radiation
in the case of $W$ production and the ensuing single-logarithms terms which cannot be simplified in the ratio. We also observe that scale variation bands at LL and at NLL+NLO do not overlap, signaling that the true perturbative uncertainty could be underestimated by correlated scale variation. A more robust perturbative uncertainty can be obtained considering also the size of
the difference between the prediction at NLL+NLO and the LL one. 

In Fig. \ref{WZqT-QCD+QED_lhc13.6_new} we show the prediction for the $R(q_T)$ distribution at the LHC.
At LL the QED corrections and the scale variation band are almost vanishing (less than per-mille level effect). Indeed, besides the simplification in the ratio of universal corrections, we also expect the suppression of quark-induced contributions with respect to the gluon-induced ones.
At NLL+NLO the QED corrections make the distribution softer at $\mathcal{O}(0.5\%)$ level. The effects vary with $q_T$, ranging from $\sim + 0.6 \%$ at $q_T \sim 2 \, \text{GeV}$ to $-0.3 \%$ at $q_T \, \sim 35\, \text{GeV}$. Scale variation band, while being quite larger than the LL one, is smaller in comparison to Tevatron case: the minimum value of about $\sim 0.1 \, \%$ is reached for $q_T \sim \, 23 \, \text{GeV}$ while the maximum one of $ 0.5 \, \%$ for $q_T \sim 40 \, \text{GeV}$. Also in this case we observe only a partial overlap of the scale uncertainty band, suggesting to also use the difference between the prediction at NLL+NLO and the LL one in order to obtain a more robust perturbative uncertainty.

\section{Conclusions}
\label{sec:Conclusions}
In this article we have combined the QED and QCD transverse-momentum ($q_T$) resummation formalisms for the production of electrically neutral and charged high-mass systems. We started from the results 
presented in Ref. \cite{Cieri:2018sfk} for on-shell $Z$ boson production obtained through an abelianization procedure of the QCD resummation framework and we extended them to the case of $W$ boson production.
However, in the $W$ boson case, a direct abelianization of QCD results is not possible, due to the presence of a charged final state and the corresponding additional (logarithmically enhanced) QED soft radiation.
\\
Therefore we performed the abelianization (along the lines of Refs.~\cite{deFlorian:2015ujt,deFlorian:2016gvk}) of the resummation formalism for a coloured final state of Ref.~\cite{Catani:2014qha}, by replacing the heavy-quark pair with a $W$ boson. As a crosscheck of our approach, in App. \ref{App:RadiativeFSR}, we performed the expansion at small $q_T$ of the real inclusive cross section, which reproduced the resummation coefficients used in our analyses. 
\\
Analytical formulas for the $q_T$ resummation of soft and collinear QED emissions were obtained at NLL for QED and LL for the mixed QCD-QED terms. The matching with fixed order predictions at leading order in QED and NLO in QCD was also performed, to properly describe the intermediate and large $q_T$ region. 
Different QED resummation effects for $Z$ and $W$ boson production appear at NLL accuracy due to soft wide-angle emission from $W$ boson in the final state.
\\
Through an implementation of our formalism in the public numerical code \texttt{DYqT}, we presented predictions at NNLL+NNLO in QCD and NLL+NLO in QED, also including the one-loop electroweak corrections both at Tevatron and LHC. We found that the resummed QED effects reach the percent level both in $Z$ and $W$ production.
The QED effects are more relevant at the Tevatron than at the LHC where the effect of the gluon parton density is larger. 
\\
Uncertainties due to missing higher-order logarithmic terms and fixed-order corrections have been obtained through combined variations of QED resummation ($Q'$) and renormalization ($\mu_R'$) scales. For on-shell $W$ boson production, 
we have found that the QED scale variation band is $\mathcal{O} (2 \% \, - \, 4\%)$ at LL, and $\mathcal{O}(1 \% - \, 1.5 \%)$ at NLL+NLO; while for on-shell $Z$ boson is $\mathcal{O} (2 \% \, - \, 4.5\%)$ at LL and $\mathcal{O}(1.5 \% - \, 3 \%)$ at NLL+NLO. In both cases, a noticeable reduction of the scale variation band is observed increasing the accuracy of our calculation. 
\\
We also considered the ratio of the $W$ and $Z$ $q_T$-distributions which is particularly relevant in the context of $m_W$ extraction. A sizeable reduction of the scale variation band is present at LL in QED, due the cancellation of common uncertainties. However, this does not happen at NLL+NLO, where contributions due to QED emission from the $W$ in the
final state cannot be simplified in the ratio of $W$ and $Z$ distributions. 
\\
A natural extension of this work is the inclusion of the decay of the weak bosons and the associated QED radiation from the charged leptons in the final state. Altogether, these corrections might allow a more precise confrontation of the theoretical predictions with the experimental data, thus leading to a more accurate determination of the EW parameters (with particular emphasis in $m_W$).

\section*{Acknowledgements}
We gratefully acknowledge Stefano Catani and Alessandro Vicini for useful discussions. The work of G.S. is partially supported by Programas Propios II (Universidad de Salamanca), EU Horizon 2020 research and innovation program STRONG-2020 project under grant agreement No. 824093 and H2020-MSCA-COFUND-2020 USAL4EXCELLENCE-PROOPI-391 project under grant agreement No 101034371. LC is supported by the Generalitat Valenciana (Spain) through the plan GenT program (CIDEGENT/2020/011) and his work is supported by the Spanish Government (Agencia Estatal de Investigación) and ERDF funds from European Commission (Grant no. PID2020-114473GB-I00 funded by MCIN/AEI/10.13039/501100011033).

\appendix
\section{Radiation from a massive final state: logarithmically-enhanced terms and linear power corrections}
\label{App:RadiativeFSR}
The logarithmic terms obtained by a fixed-order truncation of the Sudakov form factor can be equivalently found by expanding, at small $q_T$, the integrated cross section due to real photon emission  at NLO. In this way, we can check the results that we derived through the use of an abelianization-like procedure in Sec. \ref{sec:Introductionqt}. In order to perform this crosscheck, we followed the process-independent method presented in Ref. \cite{Cieri:2019tfv}, which introduces a small transverse-momentum cutoff ($q_T^{cut}$) to capture the divergent behaviour of the cross sections and the sub-leading power corrections as well.

\begin{figure} [htbp]
    \includegraphics[scale = 0.90]{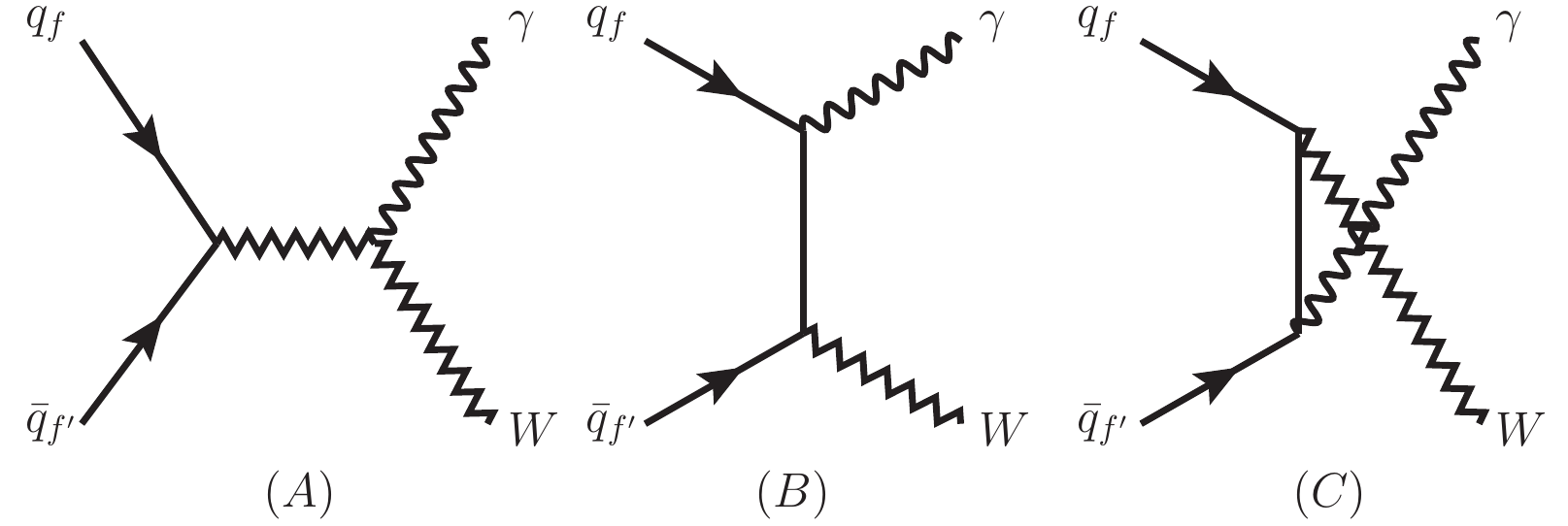}
    \centering
    \caption{On-shell $W$ boson production: photon emission at NLO in the $q_f$-$\bar{q}_{f'}$ channel.}
    \label{REAL_WONSHELL_EW1ord}
\end{figure}

We start by writing the hadronic cross section as 
\begin{equation}
 \sigma = \sum_{ab} \tau \int_{\tau}^1 \frac{d z}{z} \, \mathcal{L}_{ab} \biggl( \frac{\tau}{z} \biggl) \, \frac{1}{z} \, \int dq_T^2 \, \, \frac{d \hat{\sigma}_{ab} (q_T, z)}{d q_T^2} \, ,
\end{equation}
where $\tau = Q^2 / S$, $S$ is the hadronic squared center-of-mass energy and $Q$ is the vector boson mass. Here, $\mathcal{L}_{ab}$ is the parton luminosity and is given by 
\begin{equation}
\mathcal{L}_{ab} (y) = \int_0^{1} \frac{d x}{x} f_a (x) f_b \biggl( \frac{y}{x} \biggl).
\end{equation}
The structure at small $q_T$ of the inclusive cross section can be inferred from 
\begin{equation}
\hat{\sigma}_{ab} (z) = \int_{(q_T^{cut})^2}^{(q_T^{max})^2} dq_T^2 \, \, \frac{d \hat{\sigma}_{ab} (q_T, z)}{d q_T^2}\,,
\end{equation} 
where
\begin{equation}
(q_T^{max})^2 = Q^2 \, \frac{(1-z)^2}{4 z} \, ,
\end{equation}
is the maximum transverse momentum allowed by the kinematics of the event. Since we aim to recalculate the coefficients presented in Sec.~\ref{sec:Introductionqt}, we restrict our attention to the diagonal channel, i.e. $(a, b)=(q_{f},\bar{q}_{f'})$, which is given by the three Feynman diagrams shown in Fig. \ref{REAL_WONSHELL_EW1ord}. Then, their contribution to the cross-section is:
\begin{equation}
  \biggl(  \frac{d \hat \sigma^{(1)}}{d q_T^2} \biggl) = \biggl(  \frac{d \hat \sigma^{(1)}}{d q_T^2} \biggl)_{I.S.R.} + \biggl( \frac{d \hat \sigma^{(1)}}{ d q_T^2} \biggl)_{F.S.R.-int} \ , 
  \label{eq:EqSPLITA}
\end{equation}
where the first term in the r.h.s 
is due to the initial-state radiation, and it is given by: 
\begin{equation}
         \biggl( \frac{d \hat{\sigma}_{q \bar{q}}^{(1)}  (q_T, z)}{d q_T^2} \biggl)_{I.S.R.} = \hat{\sigma}^{(0)} \biggl( \frac{e_{q_{f}}^2+e_{\bar{q}_{f'}}^2}{2} \biggl) \,   \, z \,  \frac{-4 z \frac{q_T^2}{Q^2} + 2 (1-z) \hat{p}_{qq} (z)}{\sqrt{(1-z)^2 - 4 z \frac{q_T^2}{Q^2}}} \frac{1}{q_T^2} \, .
         \label{dec-diffXS_I.S.R.}
\end{equation}
The second term in the r.h.s. of Eq. (\ref{eq:EqSPLITA}) 
is originated by final-state radiation and its interference with the initial state, and can be written as:
\begin{equation}
    \biggl( \frac{d \hat{\sigma}_{q \bar{q}}^{(1)} (q_T, z)}{ d q_T^2} \biggl)_{F.S.R.-int}  = 
    \frac{ \hat{\sigma}^{(0)} z^2}{(1-z)^2} \frac{(-2 \hat{p}_{qq}(z) \frac{q_T^2}{Q^2}  (1-z) - 4 \frac{q_T^4}{Q^4} z)}{\sqrt{(1-z)^2- 4 z \frac{q_T^2}{Q^2}}} \frac{1}{q_T^2} \, .
    \label{dec-diffXS_calcul-real-FSR} 
\end{equation}
The functions $\hat{p}_{qq}$ are the Altarelli-Parisi unregularised splitting kernels, i.e.
\begin{equation}
\hat{p}_{qq} (z) = \frac{1+z^2}{1-z} \, . 
\end{equation}
We underline that the initial-state radiation contribution in Eq. (\ref{dec-diffXS_I.S.R.}) is obtainable from a direct abelianization of the analogous QCD process (see Eq. (2.11) of Ref. \cite{Cieri:2019tfv}) at NLO (i.e. $q \bar{q} \rightarrow W g$). Explicitly, the abelianization procedure consists in the replacement (as explained in Sec. \ref{sec:CombinedQEDQCD}): 
\begin{equation}
C_F \rightarrow \frac{e_{q_{f}} ^2 + e_{\bar{q}_{f'}}^2}{2} \, .
\end{equation}
Nonetheless, the term due to final-state radiation, i.e. Eq. (\ref{dec-diffXS_calcul-real-FSR}), has not a QCD analogous, since the $W$ boson is electrically charge but colourless. 

Having said this, we perform the integration over $q_T$, at a fixed value of $z$, obtaining
\begin{equation}
\label{tot-XS-splitted}
    \hat{\sigma}^{>(1)}  (z) = \frac{1}{C_F} \, \biggl( \frac{e_{q_{f}} ^2 + e_{\bar{q}_{f'}}^2}{2} \biggl)  \, \hat{\sigma}_{QCD}^{>(1)} (z) + \hat{\sigma}_{F.S.R. - int}^{>(1)} (z) \, ,
\end{equation}
with
\begin{equation*}
  \hat{\sigma}_{QCD}^{>(1)} (z) = \hat{\sigma}^{(0)} C_F z \biggl\{ -2 (1-z) \sqrt{1 - \frac{4 a z }{ (1-z)^{2}}}
  \end{equation*}
  \begin{equation}
  + 2 \hat{p}_{qq}(z) \biggl[ -\log \frac{a z }{ (1-z)^{2} } + 2 \log \frac{1}{2} \biggl( \sqrt{1 - \frac{4 a z}{(1-z)^{2}}} + 1 \biggl) \biggl] \biggl\} \, ,
  \label{tot-XS-QCD}
\end{equation}
and
\begin{equation}
    \hat{\sigma}_{F.S.R.- int}^{>(1)} (z) =  - \frac{\hat{\sigma}^{(0)} z \hat{p}_{qq} (z) \sqrt{(1-z)^2 - 4 a z}}{(1-z)} + \frac{\hat{\sigma}^{(0)} z \sqrt{(1-z)^2 - 4 a z} ((1-z)^2 + 2 a z )}{3 (1-z)^2} \, .
    \label{tot-XS-FSR-int}
\end{equation}
In these expressions, we introduced the dimensionless cutoff-dependent quantity
\begin{equation}
    a \equiv \frac{(q_T^{cut})^2}{Q^2} \, ,
\end{equation}
to simplify the notation. Eq. \eqref{tot-XS-QCD} is the corresponding QCD contribution to the diagonal channel for vector boson production (see Eq. (3.9) of Ref. \cite{Cieri:2019tfv}). Eq. \eqref{tot-XS-splitted} is divided in a contribution which finds its analogous in QCD (see Eq. \eqref{tot-XS-QCD}) and in a term due to the final state radiation (see Eq. \eqref{tot-XS-FSR-int}). 

The inclusive cross-section is subsequently obtained by performing the $z$ integration. 
The introduction of a transverse-momentum cutoff restricts maximum-value $z$ value that can be reached, i.e.
\begin{equation}
    z^{max} \equiv 1 - f(a), \, \, \, \, \, \, \, f(a) \equiv 2 \sqrt{a} (\sqrt{1 + a} - \sqrt{a}).
\end{equation}
This condition can be found imposing the reality of Eqs. (\ref{dec-diffXS_I.S.R.}, \ref{dec-diffXS_calcul-real-FSR}, \ref{tot-XS-QCD}, \ref{tot-XS-FSR-int}) since they represent physical quantities. 

Since we are interested on the logarithmically-enhanced terms in the small-$q_T^{cut}$ limit, we extend the integration interval up to $z \rightarrow 1$, i.e. the upper limit of $z$ in LO kinematics. This can be achieved by using the formula \cite{Cieri:2019tfv}
\begin{equation}
    \sigma_{q_{f} \bar{q}_{f'}}^{>(1)} = \tau \int_0^{1-f(a)} \frac{dz}{z} \mathcal{L}_{q_{f} \bar{q}_{f'}} \biggl( \frac{\tau}{z} \biggl) \frac{1}{z} \hat{\sigma}_{q_{f} \bar{q}_{f'}}^{(1)} (z) = \tau \int_{\tau}^{1} \frac{dz}{z} \mathcal{L}_{q_{f} \bar{q}_{f'}} \biggl( \frac{\tau}{z} \biggl) \hat{\sigma}^{(0)} \hat{G}_{q_{f} \bar{q}_{f'}}^{(1)}(z) \, .
\end{equation}
The function $\hat{G}_{q_{f} \bar{q}_{f'}}$ \cite{Cieri:2019tfv} can be expressed as a power series in the cutoff according to
\begin{equation}
   \hat{G}_{q_{f} \bar{q}_{f'}} = \sum_{m,r} 
   \log^m (a) a^{\frac{r}{2}} \hat{G}_{q_{f} \bar{q}_{f'}}^{(1,m,r)}(z). 
\end{equation}
In particular, it contains: \emph{(i)} logarithmic-enhanced cutoff dependent terms, \emph{(ii)} finite (in general process dependent) contributions and \emph{(iii)} sub-leading power corrections terms (i.e terms which vanish in the small-$q_T$ limit). In this paper, we have considered the expansion of Eq. \eqref{dec-diffXS_calcul-real-FSR} up to the first dominant sub-leading power-correction. In this case, the corresponding expression for the $\hat{G}_{q_{f} \bar{q}_{f'}}^{(1)}$ perturbative coefficient is
\begin{align}
    \hat{G}_{q_{f} \bar{q}_{f'}}^{(1)} & = \frac{1}{2} \log ^2(a) \left( \frac{e_{q_{f}} ^2 + e_{\bar{q}_{f'}}^2}{2} \right)  \, \delta (1-z) + \frac{3}{2} \log (a) 
  \left( \frac{e_{q_{f}} ^2 + e_{\bar{q}_{f'}}^2}{2} \right)  \delta (1-z)\nonumber\\
  & + \log (a) \, \frac{e^2_V}{2} \delta (1-z) - \frac{1}{2} \log (a)  \left( P^{QED}_{q_{f} q_{f}}+ P^{QED}_{\bar{q}_{f'} \bar{q}_{f'}}  \right) \nonumber\\
  & + (1+z^2) \left(- z \,\,\frac{\log (z)}{1-z} +  2 \, \left(\frac{\log (1-z)}{1-z}\right)_{+}  \right) \left( \frac{e_{q_{f}} ^2 + e_{\bar{q}_{f'}}^2}{2} \right) \nonumber\\
  & - (1+z^2)\, \, \frac{e_V^2}{2} \left( \frac{1}{1-z} \right)_{+} - (1-z) \left(   \left(\frac{e_{q_{f}} ^2 + e_{\bar{q}_{f'}}^2}{2} \right) - \frac{1}{3 \, z} \, \, \frac{e_V^2}{2} \right)\nonumber\\
  & + \left( \left(\frac{e_{q_{f}} ^2 + e_{\bar{q}_{f'}}^2}{2} \right) \frac{ \pi ^2}{6}
    + e_V^2\right) \delta (1-z)+\sqrt{a} \, \, \frac{e_V^2}{2} \left(2 \pi  \delta '(1-z)-3
   \pi  \delta (1-z)\right)\,,
   \label{Eq:G1qqexp}
\end{align}
where $P^{QED}_{q q}$ are the Altarelli-Parisi QED splitting functions given in Refs. \cite{Sborlini:2016dfn,deFlorian:2016gvk}. Eq. \eqref{Eq:G1qqexp} allows us to check explicitly (and independently of the abelianization prescription) the resummation coefficients $A'^{(1)}$ and $B'^{(1)}$ introduced in Eqs. (\ref{eq:A1})-(\ref{eq:B1}). An additional logarithmically-enhanced term proportional to $D'^{(1)} \log(a)$, $D'^{(1)} = -\frac{e_V^2}{2}$ comes from the soft emission of the $W$ boson. These expressions totally agree with the ones presented in Sec. \ref{sec:Introductionqt} (see Eq. (\ref{eq:D1})) by a direct application of the abelianization strategy to the known QCD expressions.

We also found a linear power correction in the cutoff ($\sqrt{a}$) proportional to the square of the $W$ boson charge (see last line of Eq. \eqref{Eq:G1qqexp}). This agrees with the result found in Ref. \cite{Buonocore:2019puv}, which states that the radiation from final-state massive legs gives rise to linear power corrections. 

\bibliographystyle{JHEP}


\providecommand{\href}[2]{#2}\begingroup\raggedright\endgroup

\end{document}